\documentclass[sigconf]{acmart}

\usepackage{amsmath}
\usepackage{tabularx}
\usepackage{multirow}
\usepackage{soul}
\usepackage{booktabs}
\usepackage{url}
\usepackage{siunitx}
\usepackage{subcaption}
\usepackage{xcolor}
\usepackage{adjustbox}
\usepackage{booktabs}
\usepackage{placeins}
\usepackage{csquotes}
\usepackage{bbm}
\usepackage{tcolorbox}
\usepackage{float}
\usepackage{capt-of} 

\sethlcolor{lightgray}

\AtBeginDocument{}

\copyrightyear{2026}
\acmYear{2026}
\setcctype{by}
\acmConference[WWW Companion '26]{Companion Proceedings of the ACM Web Conference 2026}{April 13--17, 2026}{Dubai, United Arab Emirates}
\acmBooktitle{Companion Proceedings of the ACM Web Conference 2026 (WWW Companion '26), April 13--17, 2026, Dubai, United Arab Emirates}
\acmDOI{10.1145/3774905.3795477}
\acmISBN{979-8-4007-2308-7/2026/04}

\begin{document}

\title{Assessing the Reliability of Persona-Conditioned LLMs as Synthetic Survey Respondents}

\titlenote{\textcolor{red}{Article published in Companion publications of \textit{WebConf'26 – 35th ACM Web Conference}. DOI: \href{http://doi.org/10.1145/3774905.3795477}{10.1145/3774905.3795477}. Please, cite the published version.}}

\author{Erika Elizabeth Taday Morocho}
\affiliation{\institution{IIT-CNR, University of Pisa}
  \city{Pisa}
  \country{Italy}}
\affiliation{
  \institution{University of Florence}
  \city{Florence}
  \country{Italy}}
\email{erika.taday@iit.cnr.it}

\author{Lorenzo Cima}
\affiliation{\institution{IIT-CNR, University of Pisa}
  \city{Pisa}
  \country{Italy}}
\email{lorenzo.cima@iit.cnr.it}

\author{Tiziano Fagni}
\affiliation{\institution{IIT-CNR}
 \city{Pisa}
 \country{Italy}}
\email{tiziano.fagni@iit.cnr.it}

\author{Marco Avvenuti}
\affiliation{\institution{University of Pisa}
  \city{Pisa}
  \country{Italy}}
\email{marco.avvenuti@unipi.it}

\author{Stefano Cresci}
\affiliation{\institution{IIT-CNR}
  \city{Pisa}
  \country{Italy}}
\email{stefano.cresci@iit.cnr.it}

\renewcommand{\shortauthors}{Taday Morocho et al.}

\begin{abstract}
Using persona-conditioned LLMs as synthetic survey respondents has become a common practice in computational social science and agent-based simulations. Yet, it remains unclear whether multi-attribute persona prompting improves LLM reliability or instead introduces distortions. Here we contribute to this assessment by leveraging a large dataset of U.S. microdata from the World Values Survey. Concretely, we evaluate two open-weight chat models and a random-guesser baseline across more than 70K respondent–item instances. We find that persona prompting does not yield a clear aggregate improvement in survey alignment and, in many cases, significantly degrades performance. Persona effects are highly heterogeneous as most items exhibit minimal change, while a small subset of questions and underrepresented subgroups experience disproportionate distortions. Our findings highlight a key adverse impact of current persona-based simulation practices: demographic conditioning can redistribute error in ways that undermine subgroup fidelity and risk misleading downstream analyses.
\end{abstract}



\maketitle

\section{Introduction}
Large language models (LLMs) are increasingly used as sources of synthetic data in survey research~\cite{argyle2023out}, in computational social science tasks~\cite{veselovsky2023generating,cima2025contextualized}, and beyond---including as synthetic survey respondents and for opinion prediction \citep{aher2023using, kim2023ai, durmus2023towards, santurkar2023whose, cao2025specializing}. In these lines of work, the LLMs are prompted to answer survey questions and their outputs are leveraged as if they were human responses, sometimes after conditioning the model on sociodemographic personas \citep{argyle2023out, alkhamissi2024investigating}. Related uses arise in social simulations and synthetic online environments, where persona-conditioned LLM agents are employed to study interaction patterns and outcomes under alternative ranking and recommendation policies \citep{tornberg2023simulating, ferraro2024agent,park2023generative}. Despite differences in application-specific objectives, both strands rely centrally on persona prompting, which motivates a careful assessment of when persona conditioning improves survey-grounded external validity and subgroup fidelity and when it may distort them. Here, we refer to \emph{external validity} to denote agreement with survey-grounded population response patterns, and \emph{subgroup fidelity} to denote agreement within demographic subgroups and preservation of between-subgroup differences \cite{bisbee2024synthetic, hu2024quantifying}. 

Despite the promise of using LLM data for social and behavioral research, other studies on LLM-generated data emphasize that LLM outputs are model-dependent artifacts rather than measurements of the world~\citep{rossi2024problems}, and that LLMs may systematically misrepresent marginalised groups if treated as observations~\citep{kozlowski2025simulating, veselovsky2023generating, moller2024parrot}. This strand of research motivates careful, use-specific evaluation of LLM outputs. This concern is particularly salient for subgroup analysis, because model bias and miscalibration may be unevenly distributed across survey items and demographic attributes. Consequently, errors can concentrate in specific groups even when aggregate agreement appears satisfactory, affecting downstream conclusions.

A growing empirical literature has therefore examined whether synthetic populations reproduce survey-grounded patterns. \citet{argyle2023out} construct ``silicon samples'' by conditioning GPT-style models on detailed sociodemographic profiles and show that, in some settings, these samples approximate subgroup response distributions, introducing the notion of algorithmic fidelity. Using survey instruments such as the World Values Survey, \citet{alkhamissi2024investigating} compare LLM and human answers under persona conditioning and document variation in alignment across demographic dimensions and culturally sensitive topics. Complementary benchmark evidence further indicates that agreement varies substantially across survey questions and demographic groups at scale \citep{santurkar2023whose}. Finally, methodological work cautions that seemingly strong agreement on aggregates may be insufficient for inference, since synthetic responses can exhibit reduced variation, altered associations, and sensitivity to prompt wording and model updates \citep{bisbee2024synthetic}. Taken together, these findings motivate further evaluations that are both more granular and more closely aligned with how personas are used in practice. In particular, \emph{multi-attribute} personas---that is, models simultaneously conditioned on multiple user attributes such as age, education, employment, and more---impose interacting constraints and can enable a model to match some attributes while distorting others, making it necessary to assess reliability jointly at the item and attribute levels. In conclusion, it is currently unclear whether multi-attribute persona prompting provides incremental gains relative to a matched no-persona control, or instead primarily redistributes errors across items and demographic subgroups \citep{bisbee2024synthetic}. 

\textbf{Contributions.} We contribute to addressing this gap through a survey-grounded, item- and attribute-level evaluation of multi-attribute persona conditioning using two open-weight chat models: \textit{(i)} \texttt{Llama-2-13B} and \textit{(ii)} \texttt{Qwen3-4B}. We focus on open-weight models to enable transparent, reproducible evaluation under realistic compute constraints~\cite{giorgi2025human}. Using the reference World Values Survey wave 7 dataset (WVS-7)~\citep{haerpfer2022world} and the question set by~\citet{alkhamissi2024investigating}, we construct each persona-question instance from an individual U.S.\ survey respondent record. Persona prompting is operationalized by adding a flexible set of declarative clauses (e.g., to condition on gender, age/education, employment status, occupation group, income level, religion, and ethnicity) before each question. 
We compare a persona-conditioned setting to \textit{(i)} an unconditioned (vanilla) baseline, which captures the model's responses under identical answer-format constraints but without socio-demographic conditioning, and \textit{(ii)} a uniform theoretical random-guesser baseline. Our goal is not only to assess overall agreement with human response patterns, but most importantly to identify the specific attributes and items for which these models appear most and least reliable---a research focus for which existing works provided limited results. In summary, we make the following main contributions:
\begin{itemize}
    \item \textbf{Controlled isolation of persona prompting.} We quantify the incremental effect of multi-attribute persona conditioning using a matched persona-based \textit{vs.} vanilla design, contextualized by a uniform random-guesser baseline.
    \item \textbf{Joint diagnostics.} We provide joint item- and attribute-level diagnostics following state-of-the-art evaluation frameworks~\cite{alkhamissi2024investigating}, including hard and soft similarity measures, to localize where models match survey-grounded patterns and where they systematically deviate.
    \item \textbf{Implications.} We discuss downstream risks for two major use cases---LLMs as synthetic survey respondents and as persona-conditioned AI agents in simulations---for which miscalibration for certain population subgroups can yield serious negative unintended consequences.
\end{itemize}

\section{Related Work}
Our work sits at the intersection of four interconnected strands of research: \textit{(i)} LLMs as synthetic survey respondents and opinion predictors; \textit{(ii)} persona conditioning, personality, and cultural or demographic alignment; \textit{(iii)} LLM-based social simulations and agent societies; and \textit{(iv)} methodological work on LLM-generated synthetic data for computational social science.

\subsection{LLMs as Synthetic Survey Respondents}
Increasingly, LLMs are used as stand-ins for human respondents and models of aggregate public opinion \cite{lin_alignsurvey_2025}. Argyle et al. \cite{argyle2023out} construct ``silicon samples'' by conditioning GPT-3 on detailed U.S.\ sociodemographic backstories and show that, for some instruments and subpopulations, synthetic samples approximate subgroup response distributions, introducing the notion of \emph{algorithmic fidelity}. Aher et al. \cite{aher2023using} propose ``Turing experiments'', in which GPT-style models simulate multiple participants in classic economic, psycholinguistic, and social-psychology studies. They demonstrate that such simulations qualitatively reproduce well-known treatment effects. Kim and Lee \cite{kim2023ai} fine-tune LLMs on repeated cross-sectional survey data and use them for opinion-prediction tasks such as retrodiction and unasked-opinion prediction, effectively filling in missing portions of time series on public attitudes. Subsequent work investigates whose opinions such models most closely reflect and how well they capture cross-cultural variation \cite{durmus2023towards,santurkar2023whose,arora2023probing}. Cao et al. \cite{cao2025specializing} specialise LLMs for simulating survey response distributions by fine-tuning on country-level results from global cultural surveys, reducing divergence between predicted and observed distributions even on unseen questions and countries. Closely related to our empirical setting, AlKhamissi et al. \cite{alkhamissi2024investigating} replicate a sociological survey in Egypt and the United States with persona-conditioned LLMs in Arabic and English, comparing model and human answers. They measure variation in alignment across demographic dimensions and culturally sensitive topics.

Collectively, these studies show that LLMs can approximate some survey marginals, but results vary across instruments, countries, and modeling choices. Importantly, few studies isolate if and how persona prompting changes item-level agreement and subgroup differences within a single standardised survey, such as the WVS. Here we contribute to filling this gap.

\subsection{LLM Personas, Personality, and Alignment}
A second strand of work examines mechanisms for steering LLM behaviour through personas and related constructs. Jiang et al. \cite{jiang2024personallm} assess whether LLM personas consistently express specified Big Five traits across personality inventories and narrative writing. Follow-up work explores architectural adaptations that encourage generated text to reflect personality, demographic, and mental-health characteristics without explicit persona prompts~\cite{la2025open}. Completing such persona and trait evaluations, Banayeeanzade et al. \cite{banayeeanzade2025psychological} introduce a phsychologically informed benchmark that evaluates steering effectiveness and trustworthiness across emotion and personality steering strategies, documenting potential side effects and behavioural shifts. Beyond trait expression, Pratelli and Petrocchi \cite{pratelli_evaluating_2025} evaluate whether Big-Five-conditioned agents reproduce personality-based variation in susceptibility to misinformation, finding partial replication alongside systematic divergences. Several other studies emphasise that persona conditioning can introduce systematic distortions. Gupta et al. \cite{gupta2023bias} show that assigning socio-demographic personas can yield deep-seated reasoning biases, with large, systematic performance differences across personas on reasoning benchmarks. Completing this evidence in interactive setting, Licato and Steinle \cite{licato_persona-infused_2025} study persona prompting in an adversarial strategic reasoning game and find that behavioural differences emerge most clearly when persona profiles are translated into structured heuristics via a mediator. Others characterise how LLM-based simulations can caricature social groups, over-emphasising stereotypical characteristics rather than realistic within-group variation~\cite{cheng2023compost}. Relatedly, work on persona-based dialog generation shows that responses can depend strongly on the interlocutor profile and may degenerate into copying biographical details in zero--shot settings~\cite{occhipinti_when_2025}.  Hu and Collier \cite{hu2024quantifying} explicitly quantify the ``persona effect'' and find that persona variables explain only a small share of variance in subjective NLP annotations, although persona prompts induce modest but statistically significant shifts in model predictions. Sun et al. \cite{sun2025sociodemographic} further show that sociodemographic prompting is not yet an effective approach for simulating subjective judgments: alignment with specific demographic groups does not consistently improve and can deteriorate for some groups. Finally, Li et al. \cite{li2025llm} show that LLM-generated personas themselves can introduce systematic biases in downstream applications such as U.S.\ election forecasting and opinion surveys, with simulated populations drifting away from real outcomes as more attributes are generated. 

Overall, this body of literature demonstrates that persona and trait-based conditioning can steer LLM outputs in meaningful ways, but it primarily evaluates internal coherence, trait expression, or coarse-grained alignment scores rather than survey-grounded external validity and subgroup fidelity at the item level, as we do.

\subsection{LLM Agents and Social Simulations}
Fully agentic social platforms are beginning to appear in practice, as illustrated by systems such as \textit{Moltbook}\footnote{\url{https://www.moltbook.com/}} and  \textit{Chirper.ai},\footnote{\url{https://chirper.ai/}} where autonomous AI agents interact in shared social spaces without direct human participation~\cite{coppolillo_harm_2026}. In parallel, a growing line of academic work embeds LLMs as agents in synthetic environments and social media digital twins \cite{gao2024large}. For example, Törnberg et al. \cite{tornberg2023simulating} initialise LLM-based social media users from survey-derived personas and simulate interactions under alternative newsfeed algorithms. Complementary work focuses on constructing population-aligned persona sets for LLM-driven social simulation, aiming to reduce distributional drift and population-level bias in the resulting synthetic societies \cite{hu_population-aligned_2025}. Relatedly, LLMs have been used to turn survey datasets into narrative personas for user research and population segmentation \cite{jung_personacraft_2025}.
Others use LLM agents with given interests and personalities in social simulations to analyse how preference-based recommendations shape homophily and echo chambers~\cite{ferraro2024agent}. Subsequent works built LLM-powered synthetic social media environments to explore content dissemination, moderation, and platform-level interventions \cite{rossetti2024social,lin2025simspark,tsirmpas2025scalable}. Beyond social media, Park et al. \cite{park2024generative} construct generative agents anchored in qualitative interviews with 1{,}052 real individuals and evaluate how well they reproduce respondents’ survey answers, personality traits, and experimental outcomes. Others treat LLMs as urban residents in personal-mobility simulations, as participants in coordination games, or as populations in which emergent social conventions and political polarisation arise \cite{wang2024large,ashery2025emergent,piao2025emergence,cai2025simulation,zhang2025llm}.

These works indicate that LLM agents can generate plausible macro-level phenomena—such as clustering, coordination, or polarisation—under different platform or policy scenarios. However, such studies often assume that synthetic LLM populations accurately reproduce the attitudes and group differences of the human populations they are meant to represent—an assumption that has not yet been convincingly validated. Here we provide new evidence that challenges this assumption and argue that systematic sanity checks of alignment and failure modes are a fundamental prerequisite for the responsible use of LLM-based synthetic populations in social simulations. 

\subsection{LLM Data for Computational Social Science}
Finally, a methodological strand evaluates LLM-generated data as input to computational social science and questions its status as ``data.'' In a case study on sarcasm detection, Veselovsky et al. \cite{veselovsky2023generating} show that naïvely generated synthetic text can be unfaithful to the real data distribution and propose grounding, filtering, and taxonomy-based strategies to improve faithfulness. Møller et al. \cite{moller2024parrot} compare human-labelled and LLM-augmented datasets across several computational social science classification tasks and document performance–cost trade-offs. Human labels tend to yield better or comparable performance, while synthetic augmentation can help on rare classes but may propagate model biases. Others explored using LLMs to evaluate a set of reliability criteria for news publishers, finding substantial agreement between an LLM and human experts on certain evaluations~\citep{pratelli2025evaluation}. More broadly, Rossi et al. \cite{rossi2024problems} argue that LLM-generated data are \emph{model-dependent artefacts} built on assumptions different from those underlying empirical measurement, and highlight risks of misrepresenting marginalised groups when synthetic outputs are treated as observations. Kozlowski and Evans \cite{kozlowski2025simulating} discuss the promise and peril of using AI stand-ins for social agents and interactions, emphasising the need to specify what such data can and cannot reveal about real societies. In a survey-research context, Bisbee et al. \cite{bisbee2024synthetic} show that LLM-generated survey responses can exhibit reduced variance, altered associations, sensitivity to prompt design, and instability across model updates, arguing that seemingly strong aggregate agreement may still be insufficient for inference. Our study is aligned with these critical perspectives as we produce concrete evidence on a survey-grounded use case: LLMs as synthetic respondents to a standardised public-opinion instrument. 

\section{Data and Methods}
\subsection{Dataset and Preprocessing}
\subsubsection{Survey data}
The data used in this study are drawn from the World Values Survey wave 7 (WVS-7) respondent-level microdata \citep{haerpfer2022world}. We restrict attention to the United States and treat observed WVS-7 responses as ground truth for evaluating synthetic answers. Following previous work~\cite{alkhamissi2024investigating}, we evaluate a curated subset of 31 WVS-7 items, as described in Appendix Section~\ref{sec:appendix-wvs}. To ensure that comparisons across prompting conditions are well-defined, each item is rendered with a consistent specification of the question text, the set of admissible response options, and the corresponding response codes. This common response specification fixes the option set presented to the models and enables uniform decoding of model outputs into WVS-7 response categories across all conditions.

\subsubsection{Persona attributes}
The unit of evaluation is a respondent--question instance, obtained by pairing a single U.S.\ respondent record with a single selected item from WVS-7. From each respondent record, we extract eight sociodemographic attributes that are used both to construct persona prompts and to define subgroup partitions useful for evaluation. These include: \textit{gender}, \textit{age}, \textit{highest educational level}, \textit{employment status}, \textit{occupational group}, \textit{income level}, \textit{religious denomination}, and \textit{ethnic group}. Prior to prompt construction and subgroup evaluation, we remove non-substantive entries via attribute-specific invalid sets  (e.g., ``don’t know'', ``no answer'', ``other''). Missing values are not imputed: when an attribute is unavailable after filtering, the corresponding persona clause is omitted from the persona description. We focus on the eight attributes above because they are standard sociodemographic dimensions in algorithmic bias and survey research~\cite{alkhamissi2024investigating,giorgi2025human}, and exhibit substantial variation in the U.S.\ WVS-7 sample. As a dispersion check, we compute normalized entropy for each attribute’s empirical marginal distribution. The resulting values indicate non-trivial heterogeneity across all selected attributes (gender: 0.691; age: 0.657; education: 0.538; occupation: 0.537; employment: 0.478; ethnicity: 0.446; religion: 0.431; income: 0.372). 

\subsection{Evaluation Setup}
We build on the survey-simulation paradigm of \cite{alkhamissi2024investigating}, which compares LLM responses, elicited under respondent personas, to human survey answer as a reference. However, our objective differs from~\cite{alkhamissi2024investigating}: rather than studying cultural alignment under changes in prompting language or pretraining mixtures, we isolate the incremental contribution of multi-attribute persona prompting itself. To do so, we benchmark persona based responses against a matched vanilla control (i.e., identical prompts without demographic clauses) and a uniform random guesser baseline, and we report both aggregate external validity and subgroup fidelity across demographic strata using WVS-7 microdata as ground truth.  We keep the entire elicitation and evaluation pipeline fixed across the two open-weight models that we consider: \texttt{Llama-2-13B} and \texttt{Qwen3-4B}.

\subsubsection{Prompting conditions and matched comparison}
For each respondent\textendash\allowbreak question instance, we elicit a model response under two matched prompting conditions:
\begin{itemize}
    \item \textit{Vanilla baseline (V)}: we use the base model simply asking to answer the questions.  
    \item \textit{Persona-based (PB)}: we personalize the answers by adding a persona description to the prompt. The description instructs the model to adopt the viewpoint of a specific respondent defined by its eight demographic attributes.
\end{itemize}
This approach allows isolating the incremental effect of demographic personalization. We chose this form of conditioning as it is the  most common practice in persona--driven use cases we analyze, while alternative conditioning options are left to future work~\cite{banayeeanzade2025psychological}. Appendix Section~\ref{sec:appendix-prompts} provides more details and examples of the prompts, items, and answers formatting. As a contextual baseline, we additionally report a theoretically uniform random-guesser (R) that samples uniformly from the valid response options for each item, enabling us to distinguish substantive agreement from chance-level matching.

\subsubsection{Decoding protocol}
Because WVS-7 items require selecting one option from a finite, explicitly enumerated list, the required generations are short. We therefore cap the continuation to 40 new tokens. Moreover,  we use a low sampling temperature of 0.3, and generate exactly one completion per respondent--item instance, which intended to reduce stochasticity. Nonetheless, we do not estimate generation variance across repeated runs, random seeds, or prompt paraphrases, and thus interpret results as point estimates under this fixed decoding regime.
We adopt this decoding regime to mirror the intended use case of single-shot survey answering and to enable matched paired persona-based \textit{vs.} vanilla comparisons. To avoid contaminating the output with prompt text, we decode only the generated continuation beyond the prompt. 

\subsubsection{Response parsing and quality control}
We map raw model answers to valid WVS-7 responses using a robust parsing routine. The parser \textit{(i)} normalizes casing and punctuation, \textit{(ii)} removes leading enumeration artifacts, and \textit{(iii)} resolves responses by matching either an explicit option index or a text span that uniquely identifies a candidate option string.
The parsing procedure is applied identically under persona-conditioned and vanilla prompting to preserve the matched comparison. Because we enforce strict answer formatting, the number of retained personas can differ slightly across models due to occasional unparseable outputs. Tables~\ref{tab:overall-results}--\ref{tab:results-attributes} report the resulting valid counts per condition under the \textit{users} column.

\subsubsection{Evaluation metrics}
Let $\mathcal{Q}$ be the set of evaluated survey items and $\mathcal{P}$ the set of personas (i.e., respondent profiles). For each item $q\in\mathcal{Q}$ and persona $p\in\mathcal{P}$, let $y_{q,p}$ denote the ground-truth WVS-7 response code and $\hat{y}_{q,p}^{\,c}$ the decoded model response under condition $c$ (persona-based \textit{vs.} vanilla). We compute two complementary agreement metrics as in the following.

\textbf{Hard similarity (HS)}, corresponding to the model \textit{accuracy}.
\begin{equation}
H_{c} \;=\; \operatorname{mean}_{q,p}\left\{\mathbbm{1}\!\left(\hat{y}_{q,p}^{\,c}=y_{q,p}\right)\right\}.
\end{equation}

\textbf{Soft similarity (SS).} Since 22 over 31 survey items are ordinal, we introduce an ordinal evaluation metric that captures the degree of disagreement beyond binary accuracy. Let $|q|$ be the number of valid response options for item $q$. We define an item-wise error term based on the Normalized Match Distance (NMD):
\begin{equation}
\varepsilon_{c}(q,p) \;=\;
\begin{cases}
\displaystyle \frac{\left|\hat{y}_{q,p}^{\,c}-y_{q,p}\right|}{|q|-1} & \text{if } \mathrm{IsOrd}(q,p),\\[8pt]
\mathbbm{1}\!\left(\hat{y}_{q,p}^{\,c}\neq y_{q,p}\right) & \text{otherwise},
\end{cases}
\end{equation}
where $\mathrm{IsOrd}(q,p)$ indicates that $q$ is treated as ordinal (e.g., Likert items). For non-ordinal items—and for cases involving non-substantive options such as \emph{Don't know}—soft similarity reduces to a 0/1 mismatch, aligning it with hard similarity on those instances. We then define SS as:
\begin{equation}
S_{c} \;=\; \operatorname{mean}_{q,p}\left\{1-\varepsilon_{c}(q,p)\right\}.
\end{equation}

\begin{table}[t]
    \small
    \setlength{\tabcolsep}{1.4pt}
	\centering
	\begin{tabular}{lrrrrrccccccc}
        \toprule
        & \multicolumn{2}{c}{\multirow{2}{*}{\textbf{vanilla (V)}}} && \multicolumn{3}{c}{\textbf{persona}} && \multicolumn{2}{c}{\multirow{2}{*}{\textbf{PB \textit{vs.} R}}} && \multicolumn{2}{c}{\multirow{2}{*}{\textbf{PB \textit{vs.} V}}} \\
		&&&& \multicolumn{3}{c}{\textbf{based (PB)}} &&&&&  \\
		\cmidrule{2-3} \cmidrule{5-7}  \cmidrule{9-10} \cmidrule{12-13}
        \textbf{model} & \textit{HS} & \textit{SS} && \textit{users} & \textit{HS} & \textit{SS} && \textit{$p_{HS}$} & \textit{$p_{SS}$} && \textit{$p_{HS}$} & \textit{$p_{SS}$} \\
        \midrule
        random guesser (R) & 0.273 & 0.537 && 2,596 & -- & -- && -- & -- && -- & -- \\
        \texttt{Llama-2-13B} & 0.370 & 0.621 && 2,345 & 0.366 & 0.612 && ** & ***  \\
        \texttt{Qwen3-4B} & 0.391 & \textbf{0.627} && 2,359 & \textbf{0.398} & \textbf{0.627} && *** & *** \\
		\bottomrule
        \multicolumn{8}{l} {*: $p < 0.1$; **: $p < 0.05$; ***: $p < 0.01$}
	\end{tabular}
	\caption{Overall hard and soft similarity (HS and SS) scores for the persona-based model (PB), the vanilla model (V), and the random guesser (R). The users column reports the number of personas for which valid answers were obtained. Wilcoxon statistical significance for the PB \textit{vs.} R and PB \textit{vs.} V comparisons is also reported. The highest score for each metric is highlighted in bold.}
\label{tab:overall-results}
\end{table}
 
\textbf{Subgroup fidelity.} For any demographic subgroup $g\subseteq\mathcal{P}$ (e.g., sex, age, education), we compute the HS and SS metrics restricted to $p\in g$, yielding $H_{c}(g)$, $S_{c}(g)$, to compare how agreement varies across demographics slices. Subgroup membership is defined from the WVS-7 respondent attributes and is used to slice both PB and V predictions.

\section{Results and Discussion}
We evaluate PB against V, using the WVS-7 U.S. respondent-level microdata and the curated set of 31 items. Since the two conditions differ only in the inclusion of demographic clauses, PB--V differences can be interpreted as the incremental effect of demographic conditioning. We report results in terms of HS and SS, and we contextualize them in comparison to a uniform random-guesser baseline (R).

\textbf{Overall external validity.}
Table~\ref{tab:overall-results} summarizes aggregate agreement with WVS-7 responses. 
The random-guesser provides a chance baseline ($HS=0.273$, $SS=0.537$). Both V and PB variants substantially outperform R, with \texttt{Qwen3-4B} achieving higher overall scores than \texttt{Llama-2-13B}. However, PB prompting does not yield a consistent aggregate improvement across models, as also reflected in the lack of statistically significant differences between PB and V. For \texttt{Llama-2-13B}, PB slightly reduces both HS and SS relative to V, whereas for \texttt{Qwen3-4B}, PB yields a modest increase in HS while leaving SS unchanged. These aggregate results suggest that the marginal effect of persona prompting is model-dependent, and that improvements are not uniform across metrics.

\begin{figure*}[t]
  \centering
  \begin{minipage}[t]{0.265\linewidth}
    \centering
    \includegraphics[width=\columnwidth]{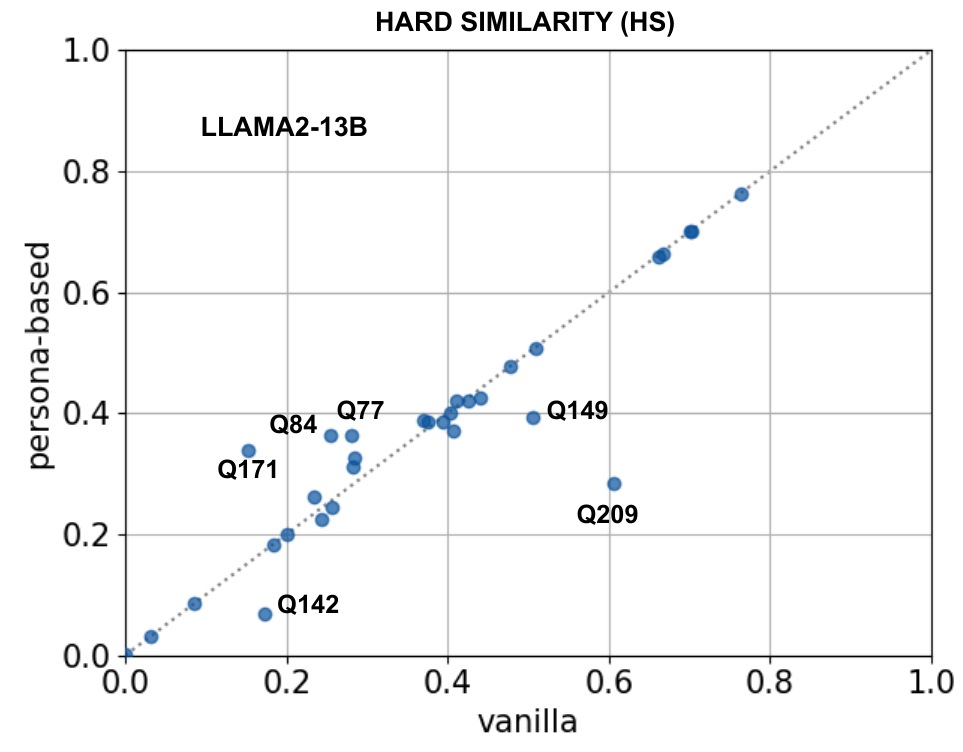}
  \end{minipage}\begin{minipage}[t]{0.265\linewidth}
    \centering
    \includegraphics[width=\columnwidth]{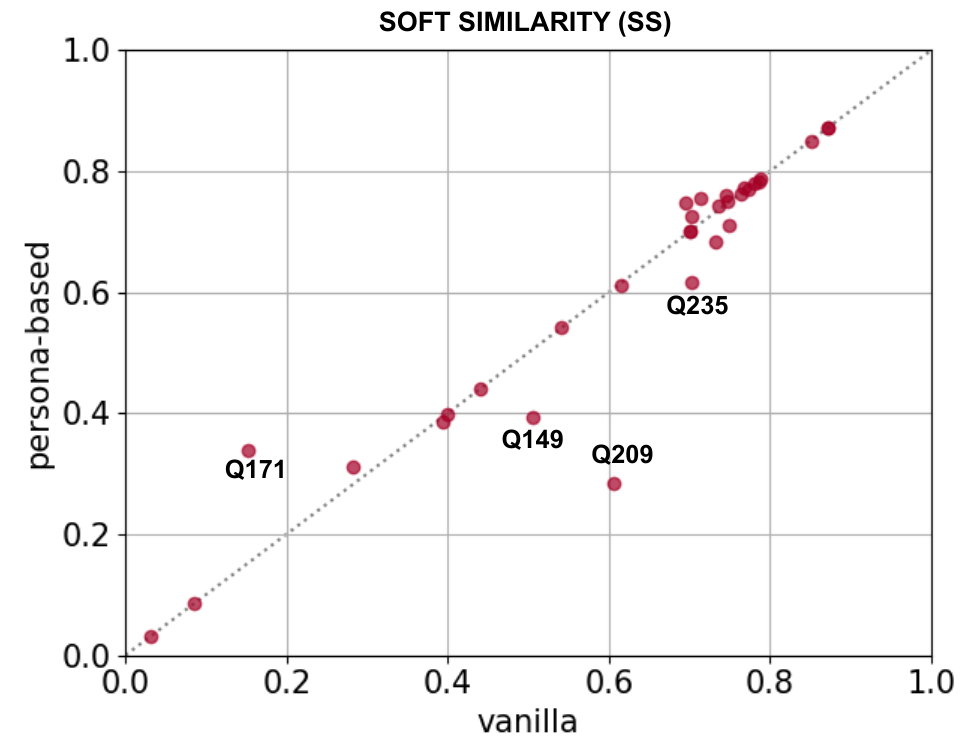}
  \end{minipage}\begin{minipage}[t]{0.47\linewidth}
    \centering
    \includegraphics[width=\columnwidth]{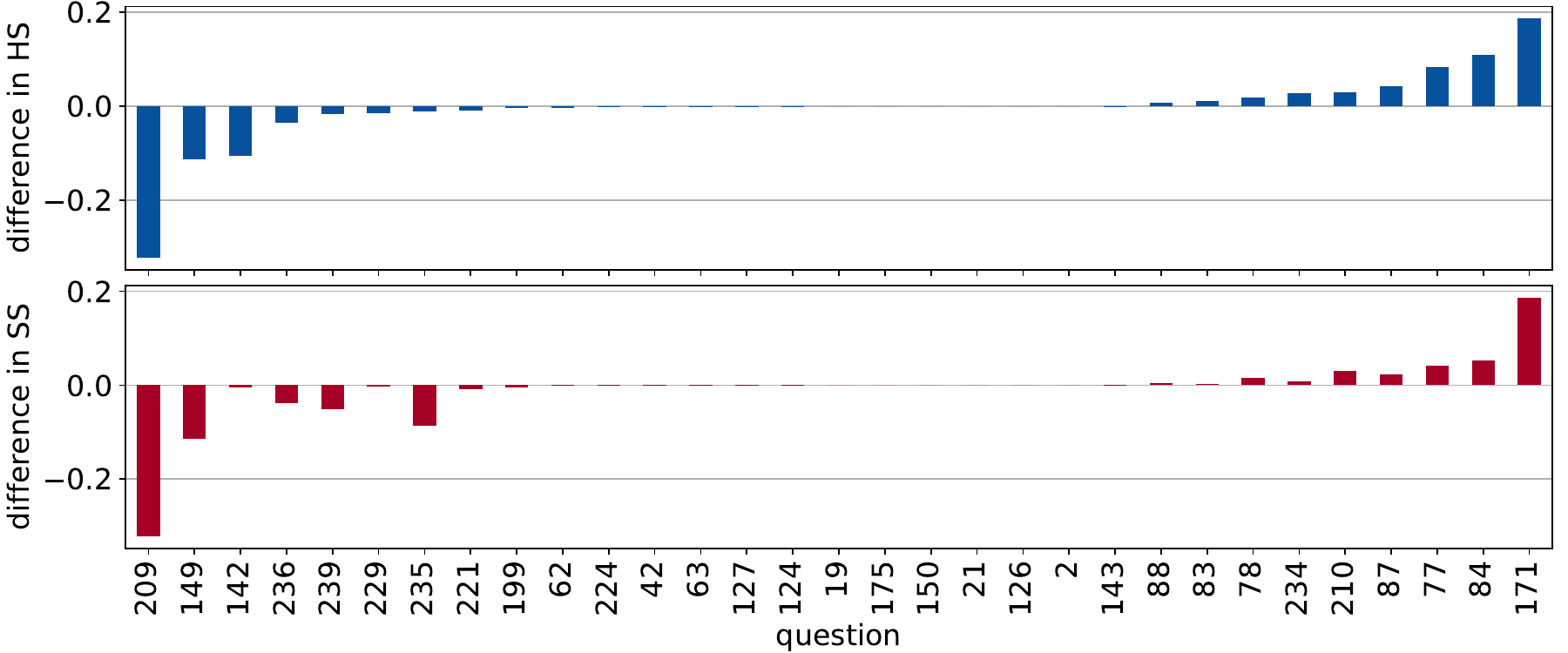}
  \end{minipage}
\begin{minipage}[t]{0.265\linewidth}
    \centering
    \includegraphics[width=\columnwidth]{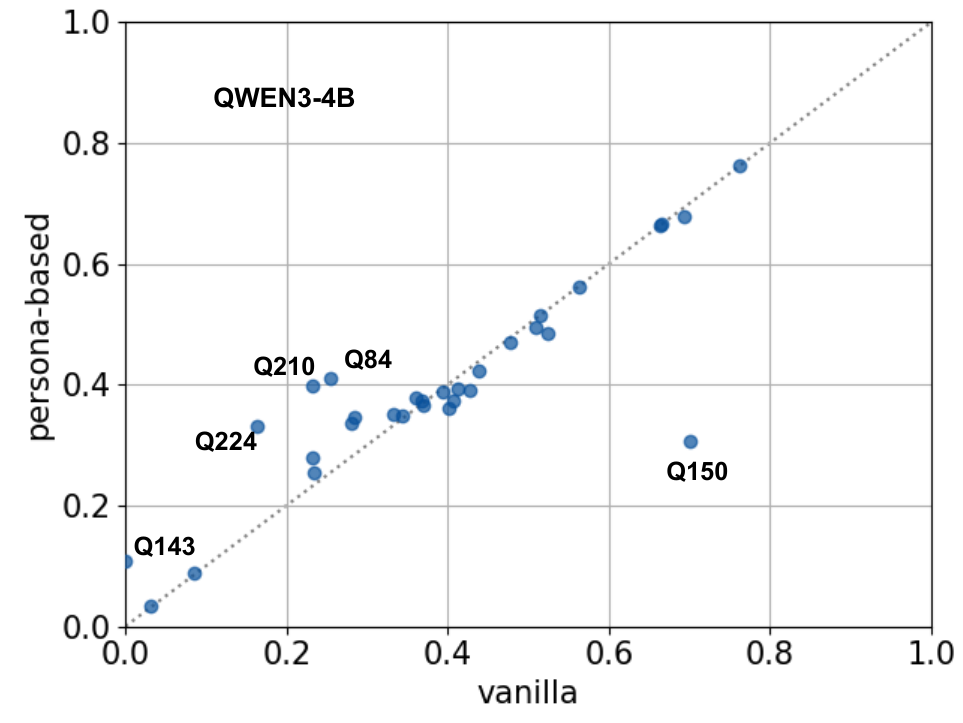}
  \end{minipage}\begin{minipage}[t]{0.265\linewidth}
    \centering
    \includegraphics[width=\columnwidth]{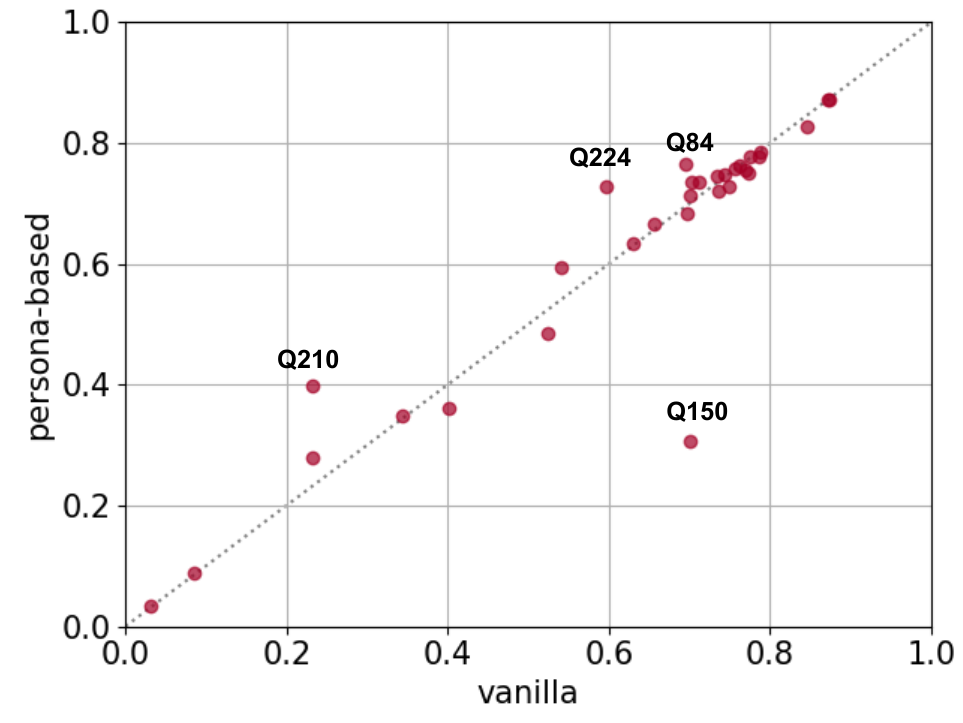}
  \end{minipage}\begin{minipage}[t]{0.47\linewidth}
    \centering
    \includegraphics[width=\columnwidth]{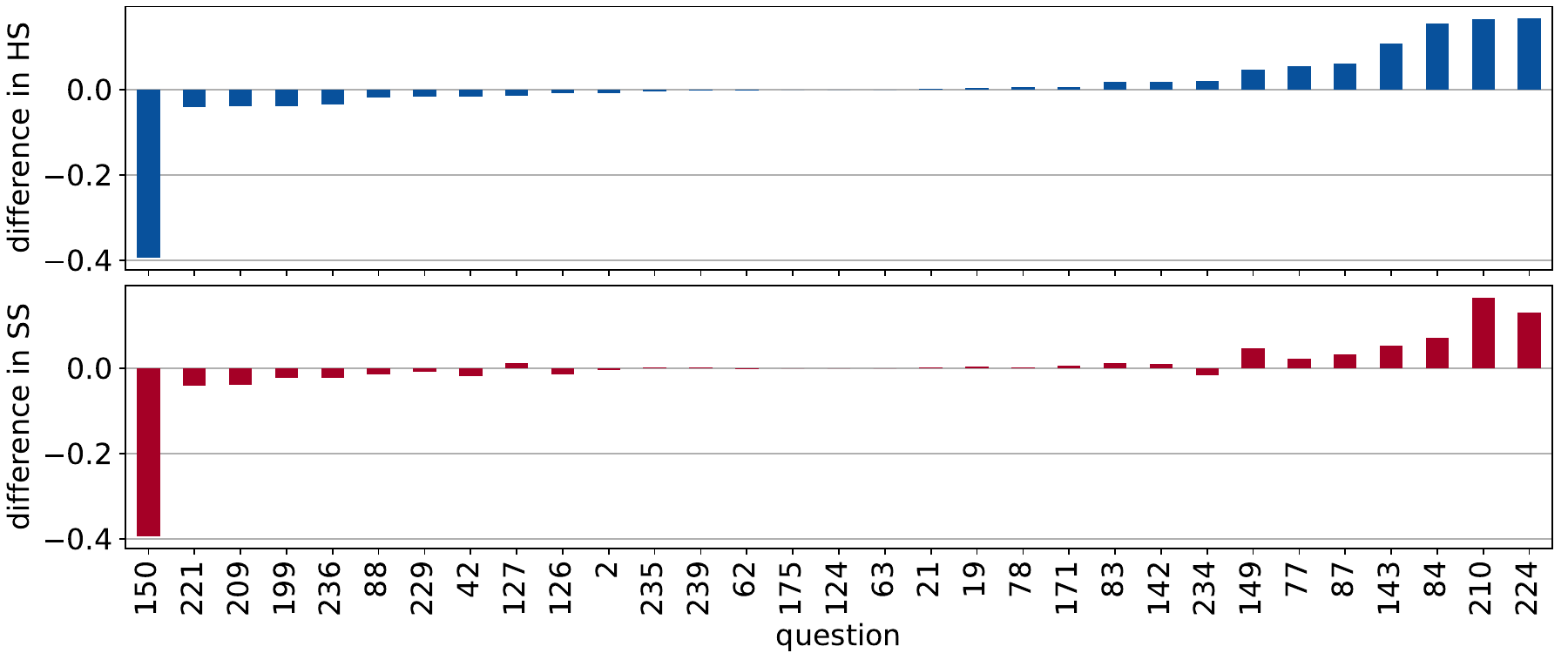}
  \end{minipage}
  \caption{Detailed results for the \texttt{Llama-2-13B} (top row) and \texttt{Qwen3-4B} (bottom row) models. For each model, we report a comparison of hard similarity (HS, blue-colored, left panel) and soft similarity (SS, red-colored, central panel) scores obtained for each question using the vanilla (V, \textit{x}-axis) and persona-based (PB, \textit{y}-axis) versions of the models. Additionally, for each model, the right panel shows the item-wise differences between the PB and V model variants, in terms of HS (blue-colored, top) and SS (red-colored, bottom). Positive differences ($>0$) indicate that PB outperforms V for the corresponding item and metric.}
  \label{fig:detailed-results}
\end{figure*}

\textbf{Item-level persona effects.}
Figure~\ref{fig:detailed-results} shows the item-level comparison between the persona-based (PB) and vanilla (V) variants of each model. Each row corresponds to a model, while columns report, from left to right: \textit{(i)} scatterplots of hard similarity (HS, blue-colored) scores obtained under V (\textit{x}-axis) and PB (\textit{y}-axis) for each survey item; \textit{(ii)} the analogous comparison for soft similarity (SS, red-colored); and \textit{(iii)} bar charts of the item-wise PB–V differences for both HS (blue-colored, top) and SS (red-colored, bottom). Points close to the diagonal in the scatterplots and bars near zero in the bar charts indicate similar behavior under PB and V, whereas larger deviations highlight items for which persona prompting substantially shifts model performance.

As depicted in Figure~\ref{fig:detailed-results}, for both models, item-wise scores under PB and V largely track each other. Most points lie close to the diagonal in the HS and SS scatter plots, and most PB--V differences cluster near zero in the bar charts. However, a limited set of WVS-7 items accounts for the largest swings in both HS and SS, indicating that persona prompting mainly acts as a selective perturbation rather than a uniform improvement. This effect is also model-dependent in magnitude, as \texttt{Qwen3-4B} exhibits larger item-level swings than \texttt{Llama-2-13B}, consistent with \texttt{Qwen3-4B} being more responsive to persona conditioning in this setup. A notable example is the ordinal item Q83 (confidence in the United Nations). On this item, PB improves HS for both models, indicating that persona clauses more often steer the model toward the exact WVS-7 response category. In SS, however, the improvement is visible for \texttt{Qwen3-4B} but not for \texttt{Llama-2-13B}. Because SS combines exact matches and ordinal distance, \texttt{Qwen3-4B}'s SS gain on Q83 indicates a lower average ordinal distance overall. On the contrary, \texttt{Llama-2-13B}'s unchanged SS despite higher HS suggests that any HS gains are offset by unchanged or larger ordinal distances on the remaining mismatches. This illustrates why reporting only exact matches can be insufficient for interpreting reliability on ordinal attitude questions.

\begin{table*}[ht!]
\centering
\small
\setlength{\tabcolsep}{2.3pt}
\scalebox{0.95}{\begin{tabular}{llllcllcrrrcllcrrr}
\toprule
&&&&& \multicolumn{6}{c}{\texttt{Llama-2-13B}} && \multicolumn{6}{c}{\texttt{Qwen3-4B}} \\
\cmidrule{6-11} \cmidrule{13-18}
&& \multicolumn{2}{c}{\textbf{random guesser (R)}} && \multicolumn{2}{c}{\textbf{vanilla (V)}} && \multicolumn{3}{c}{\textbf{persona-based (PB)}} && \multicolumn{2}{c}{\textbf{vanilla (V)}} && \multicolumn{3}{c}{\textbf{persona-based (PB)}} \\
\cmidrule{3-4} \cmidrule{6-7} \cmidrule{9-11} \cmidrule{13-14} \cmidrule{16-18}
\textbf{attribute} & \textbf{value} & \textit{HS} & \textit{SS} && \textit{HS} & \textit{SS} && \textit{users} & \textit{HS} & \textit{SS} && \textit{HS} & \textit{SS} && \textit{users} & \textit{HS} & \textit{SS} \\
\midrule
\multirow{6}{*}{age group} & $<20$ & 0.281*** & 0.547*** && \underline{0.396} & \underline{0.634} && 61 & \underline{0.390} & \underline{0.629} && \underline{0.418} & \underline{0.635} && 61 & \underline{0.414} & \underline{0.634} \\
& 20-29 & 0.273*** & 0.536*** && 0.365 & 0.620 && 512 & \hl{0.370} & \underline{\hl{0.621}} && 0.386 & 0.624 && 514 & \hl{0.398} & \hl{0.631} \\
& 30-39 & 0.280*** & 0.543*** && 0.368 & 0.619 && 549 & 0.365 & 0.614 && 0.388 & 0.626 && 551 & \hl{0.398} & \hl{0.629} \\
& 40-49 & 0.276*** & 0.537*** && 0.368 & 0.618 && 423 & 0.365 & 0.610 && 0.387 & 0.627 && 425 & \hl{0.391} & 0.624 \\
& 50-59 & 0.278*** & 0.540*** && 0.374 & 0.621 && 340 & 0.360 & 0.602 && 0.397 & 0.629 && 341 & 0.394 & 0.621 \\
& 60+ & 0.273*** & 0.528*** && 0.374 & 0.623 && 468 & 0.367 & 0.606 && 0.396 & 0.630 && 471 & \underline{\hl{0.405}} & 0.626 \\
\midrule
\multirow{3}{*}{education} & lower & 0.274** & 0.537** && 0.329 & 0.590 && 45 & \hl{0.362} & \hl{0.610} && 0.337 & 0.584 && 45 & \hl{0.361} & \hl{0.601} \\
& middle & 0.276*** & 0.540*** && 0.358 & 0.611 && 1,127 & 0.358 & 0.605 && 0.371 & 0.615 && 1,136 & \hl{0.385} & \hl{0.619} \\
& higher & 0.276*** & 0.534*** && \underline{0.384} & \underline{0.632} && 1,148 & \underline{0.375} & \underline{0.618} && \underline{0.413} & \underline{0.642} && 1,152 & \underline{0.413} & 0.637 \\
\midrule
\multirow{7}{*}{employment status} & student & 0.278*** & 0.541*** && \underline{0.407} & \textbf{0.650} && 29 & \textbf{\hl{0.412}} & \textbf{\hl{0.657}} && \textbf{0.448} & \textbf{0.668} && 29 & \underline{0.433} & \textbf{0.655} \\
& homemaker & 0.274*** & 0.540*** && 0.362 & 0.621 && 125 & 0.357 & 0.610 && 0.378 & 0.624 && 169 & \hl{0.388} & 0.616 \\
& unemployed & 0.277*** & 0.541*** && 0.347 & 0.605 && 181 & \hl{0.365} & \hl{0.613} && 0.369 & 0.611 && 181 & \hl{0.392} & \hl{0.616} \\
& self employed & 0.274*** & 0.535*** && 0.358 & 0.614 && 133 & \hl{0.366} & 0.601 && 0.398 & \underline{0.634} && 133 & 0.394 & 0.618 \\
& part time & 0.274*** & 0.539*** && 0.377 & 0.624 && 266 & 0.370 & \underline{0.617} && \underline{0.402} & 0.632 && 267 & \underline{\hl{0.412}} & \underline{\hl{0.637}} \\
& full time & 0.277*** & 0.539*** && 0.376 & 0.625 && 1,173 & 0.367 & 0.614 && 0.394 & 0.630 && 1,176 & \hl{0.396} & 0.630 \\
& retired/pensioned & 0.272*** & 0.526*** && 0.368 & 0.617 && 308 & 0.365 & 0.603 && 0.387 & 0.622 && 308 & \underline{\hl{0.408}} & \hl{0.625} \\
\midrule
\multirow{4}{*}{ethnicity} & black & 0.283*** & 0.546*** && 0.337 & 0.600 && 206 & \hl{0.339} & 0.598 && 0.352 & 0.600 && 206 & \hl{0.368} & \hl{0.604} \\
& hispanic & 0.275*** & 0.539*** && 0.355 & 0.613 && 446 & \hl{0.357} & 0.611 && 0.371 & 0.613 && 447 & \hl{0.381} & \hl{0.619} \\
& white & 0.275*** & 0.535*** && \underline{0.380} & 0.626 && 1,495 & \underline{0.373} & 0.614 && \underline{0.403} & \underline{0.636} && 1,502 & \underline{\hl{0.407}} & 0.632 \\
& two plus & 0.289*** & 0.540*** && 0.356 & 0.615 && 84 & \hl{0.357} & 0.609 && 0.389 & 0.629 && 84 & \hl{0.398} & 0.628 \\
\midrule
\multirow{3}{*}{income level} & low & 0.275*** & 0.536*** && 0.347 & 0.602 && 506 & \hl{0.366} & \hl{0.611} && 0.369 & 0.615 && 509 & \hl{0.385} & \hl{0.616} \\
& medium & 0.277*** & 0.540*** && 0.379 & \underline{0.627} && 1,595 & 0.367 & 0.614 && 0.398 & 0.632 && 1,600 & \hl{0.403} & \underline{\hl{0.633}} \\
& high & 0.271*** & 0.523*** && 0.371 & 0.622 && 199 & \underline{\hl{0.373}} & 0.605 && 0.400 & 0.631 && 199 & \underline{\hl{0.405}} & 0.615 \\
\midrule
\multirow{12}{*}{occupation group} & clerical & 0.277*** & 0.542*** && \underline{0.380} & \underline{0.632} && 166 & \underline{0.373} & \underline{0.616} && 0.399 & \underline{0.636} && 167 & \hl{0.403} & \underline{0.636} \\
& farm owner & 0.242* & 0.521** && \textbf{0.411} & 0.607 && 4 & \underline{0.387} & 0.591 && 0.336* & 0.577** && 4 & \hl{0.395} & \underline{\hl{0.637}} \\
& farm worker & 0.273* & 0.544** && 0.353 & 0.612 && 15 & 0.345 & 0.595 && 0.343 & 0.600 && 15 & \hl{0.364} & \hl{0.604} \\
& higher administrative & 0.278*** & 0.540*** && \underline{0.383} & \underline{0.631} && 97 & \underline{0.375} & \underline{0.623} && \underline{0.401} & 0.633 && 97 & \underline{\hl{0.406}} & \underline{\hl{0.637}} \\
& inap & 0.276*** & 0.535*** && 0.362 & 0.614 && 775 & \hl{0.365} & 0.609 && 0.383 & 0.622 && 777 & \hl{0.399} & 0.621 \\
& professional and technical & 0.272*** & 0.533*** && \underline{0.386} & \underline{0.633*} && 455 & \underline{0.373} & \underline{0.616} && \underline{0.416} & \underline{0.642} && 457 & \underline{\hl{0.418}} & \underline{0.641} \\
& sales & 0.284*** & 0.549*** && 0.362 & 0.618 && 153 & 0.354 & 0.607 && 0.395 & \underline{0.634} && 153 & 0.395 & \underline{0.633} \\
& semi-skilled worker & 0.274*** & 0.537*** && 0.368 & 0.619 && 67 & 0.362 & 0.606 && 0.388 & 0.624 && 67 & \hl{0.393} & 0.620 \\
& service & 0.282*** & 0.539*** && 0.372 & 0.619 && 181 & 0.372 & \underline{0.616} && 0.394 & 0.630 && 181 & \hl{0.400} & 0.630 \\
& skilled worker & 0.275*** & 0.542*** && 0.367 & 0.615 && 134 & 0.360 & 0.608 && 0.377 & 0.619 && 134 & 0.375 & 0.618 \\
& unskilled worker & 0.283** & 0.535*** && 0.357 & 0.610 && 79 & \hl{0.362} & 0.608 && 0.391 & 0.630 && 79 & 0.374 & 0.610 \\
& never had a job & 0.267* & 0.512 && \underline{0.392**} & \underline{0.628***} && 13 & 0.306 & 0.562 && 0.381 & 0.611 && 12 & 0.356 & 0.573 \\
\midrule
\multirow{9}{*}{religion} & buddhist & 0.257*** & 0.515*** && 0.347 & 0.607 && 27 & \hl{0.360} & \hl{0.610} && \underline{0.409} & \underline{0.644} && 27 & 0.383 & 0.610 \\
& catholic & 0.280*** & 0.541*** && \underline{0.380} & \underline{0.627} && 530 & 0.372 & 0.615 && 0.388 & 0.624 && 532 & \hl{0.395} & \hl{0.625} \\
& hindu & 0.267*** & 0.511*** && \underline{0.414} & \underline{0.635} && 14 & \underline{0.396} & \underline{0.633} && \underline{0.412} & 0.617 && 14 & \textbf{\hl{0.429}} & \underline{\hl{0.637}} \\
& jew & 0.264*** & 0.521*** && 0.373 & 0.626** && 46 & 0.368 & 0.601 && \underline{0.404} & \underline{0.638} && 46 & 0.398 & 0.618 \\
& muslim & 0.293* & 0.547** && 0.363 & 0.617 && 16 & 0.356 & 0.610 && 0.350 & 0.601 && 16 & \hl{0.386} & \hl{0.612} \\
& orthodox & 0.284 & 0.554* && 0.308 & 0.586 && 15 & \hl{0.319} & \hl{0.591} && 0.328 & 0.585 && 15 & \hl{0.371} & \hl{0.612} \\
& other christian & 0.283** & 0.552** && 0.363 & 0.616 && 95 & 0.340 & 0.592 && 0.371 & 0.615 && 95 & \hl{0.385} & \hl{0.616} \\
& protestant & 0.275*** & 0.533*** && \underline{0.383} & 0.626 && 451 & 0.369 & 0.610 && 0.393 & 0.624 && 452 & \hl{0.395} & 0.622 \\
& atheist & 0.275*** & 0.538*** && 0.363 & 0.617 && 1,068 & \hl{0.367} & 0.614 && 0.396 & \underline{0.634} && 1,072 & \underline{\hl{0.405}} & \underline{\hl{0.635}} \\
\midrule
\multirow{2}{*}{gender} & female & 0.277*** & 0.541*** && 0.371 & 0.622 && 1,101 & 0.366 & 0.612 && 0.386 & 0.624 && 1,103 & \hl{0.402} & \hl{0.630} \\
& male & 0.275*** & 0.534*** && 0.369 & 0.619 && 1,249 & 0.366 & 0.611 && 0.396 & 0.630 && 1,256 & 0.395 & 0.625 \\
\bottomrule
\multicolumn{18}{l} {*: $p < 0.1$; **: $p < 0.05$; ***: $p < 0.01$}
\end{tabular}}
\caption{Hard and soft similarity (HS and SS) scores for the persona-based models (PB), vanilla models (V), and the random guesser (R), aggregated by socio-demographic attribute values. Uninformative values (e.g., “do not know,” “no answer,” or “other”) are omitted. Wilcoxon statistical significance is reported for the differences between PB and V (shown in the V columns) and between the best-performing PB model and R (shown in the R columns). The highest and top-10 scores for each model and metric are highlighted in bold and underlined, respectively. Cells corresponding to evaluation scores for which the PB model outperforms the corresponding V baseline are highlighted in gray.}
\label{tab:results-attributes}
\end{table*}
 
\textbf{Subgroup fidelity and small-group instability.}
Table \ref{tab:results-attributes} reports agreement within demographic subgroups defined by the persona attributes. Two qualitative patterns emerge. First, subgroup scores under PB do not show a consistent improvement over V. In Table \ref{tab:results-attributes}, this condition is reflected with gray-highlighted cells (i.e., gray indicates PB exceeds V). Indeed, for some strata, PB slightly outperforms V, while for others it does not. This mixed behaviour is present for both models, though \texttt{Qwen3-4B} more often matches or slightly exceeds its vanilla baseline than \texttt{Llama-2-13B} does. Second, the largest PB-induced shifts tend to occur in small-\emph{n} strata (i.e., lower number of users). There, estimates are inherently higher-variance and persona prompting can introduce disproportionate shifts. This highlights a practical risk for persona-based simulation: even when aggregate agreement appears stable, demographic conditioning can redistribute errors unevenly across groups, undermining subgroup fidelity and potentially distorting downstream analyses that rely on subgroup comparisons or simulated agent populations.

Overall, across our WVS-based evaluation, multi-attribute persona prompting does not deliver a consistent aggregate gain over a matched vanilla control across models, and its effects are concentrated in a small subset of items and in low-support subgroups. These results support treating persona prompting as an intervention whose reliability is \textit{item-} and \textit{subgroup-dependent}, motivating fine-grained audits beyond aggregate scores when LLMs are used as synthetic survey respondents or as persona-conditioned agents in social simulations. 

\textbf{Limitations.} Our findings should be interpreted in light of several limitations. First, the evaluation is restricted to WVS-7 respondents from the U.S. and to a curated subset of survey items. While broadly applicable, our results may thus not transfer to other countries or questions with different cultural contexts and response distributions. Second, while we evaluate two open-weight chat models under a fixed pipeline, results may vary across additional checkpoints, prompt templates, or decoding choices.
Third, our soft metric relies on an ordinality assumption for scale items. While appropriate for Likert-style questions, it is a coarse approximation for items whose semantics are not strictly linear in option index. Finally, persona prompting is derived from a limited set of socio-demographic attributes and does not capture the full context behind a respondent’s beliefs. As a result, personas may sometimes act as stereotypes or weak proxies rather than faithful individual-level conditioning. Moreover, survey answers themselves are not error-free (e.g., measurement noise, social desirability, respondent inconsistency), and we do not model the WVS-7 sampling design or weights, which can matter when interpreting population-level representativeness.

\section{Conclusions}
Motivated by the widespread adoption of LLMs as synthetic survey respondents and agents in social simulations, we studied whether multi-attribute persona prompting improves the reliability of LLM answers. We followed a matched design where persona-based and vanilla prompts differ only by the inclusion of demographic clauses. Using World Value Survey U.S. microdata as ground truth, we evaluated two open-weight models (i.e., \texttt{Llama-2-13B} and \texttt{Qwen3-4B}) with both \textit{hard} and \textit{soft} similarity metrics, contextualized against a uniform random-guesser baseline. We found that both models clearly exceed random performance, and that \texttt{Qwen3-4B} shows higher overall survey-grounded agreement than \texttt{Llama-2-13B}. However, persona prompting did not produce a consistent aggregate gain across models. More importantly, persona effects are heterogeneous: most items exhibited limited changes, while a smaller subset exhibited larger shifts. Moreover, subgroup fidelity did not improve consistently across demographic partitions, sometimes degrading noticeably in low-support strata. Overall, these results argue against treating persona prompting as uniformly beneficial, and instead motivate viewing it as an intervention whose impact depends on the item and the population subgroup.

For practical deployments (e.g., synthetic polling, agent initialization, or subgroup analysis), we recommend reporting matched vanilla baselines and auditing item- and subgroup-level behavior, instead of only computing aggregate scores. Future work should broaden the set of considered models and the geographic scope, incorporate repeated runs to quantify variance, test additional prompt designs and languages, and adopt evaluation procedures that better respect survey design and distributional targets.

\begin{acks}
This work was supported by the European Union -- Next Generation EU, Mission 4 Component 1, project PIANO (CUP B53D23013290006); by the ERC project DEDUCE under grant \#101113826; by the PNRR-M4C2 (PE00000013) “FAIR-Future Artificial Intelligence Research" - Spoke 1 "Human-centered AI", funded under Next Generation EU; and by the Italian Ministry of Education and Research (MUR) in the framework of the FoReLab projects (Departments of Excellence).
\end{acks}

\bibliographystyle{ACM-Reference-Format}
\bibliography{references}

\appendix

\begin{table*}[t]
    \small
    \setlength{\tabcolsep}{8pt}
    \begin{tabular}{@{}lp{0.86\textwidth}@{}}
        \toprule
        \textbf{id} & \textbf{question}\\
        \midrule
        Q2   & How important are friends in life?\\
        Q19  & Do you think it is important to have neighbors who are people of a different race?\\
        Q21  & How important do you think it is to have neighbors who are immigrants/foreign workers?\\
        Q42  & What are your thoughts on the overall way our society is organized?\\
        Q62  & How much do you trust people of another religion?\\
        Q63  & How much do you trust people of another nationality?\\
        Q77  & How much confidence do you have in major companies?\\
        Q78  & Please rate your confidence level in private banks.\\
        Q83  & How much confidence do you have in the United Nations (UN)?\\
        Q84  & How much confidence do you have in the International Monetary Found (IMF)?\\
        Q87  & How much confidence do you have in the World Bank (WB)?\\
        Q88  & How much confidence do you have in the World Health Organization (WHO)?\\
        Q124 & Does immigration in your country increase the crime rate?\\
        Q126 & Do you agree that immigration in your country increases the risks of terrorism?\\
        Q127 & Do you think immigration in your country helps poor people establish new lives?\\
        Q142 & How worried are you about losing your job or not finding a job? \\ Q143 & To what extent are you worried about not being able to give your children a good education?\\
        Q149 & In your opinion, is freedom or equality more important?\\
        Q150 & Which is more important - freedom or security?\\
        Q171 & How frequently do you attend religious services?\\
        Q175 & What is the purpose of religion? \\ Q199 & How interested are you in politics?\\
        Q209 & Have you ever signed a political action petition?\\
        Q210 & Have you ever participated in a political boycott?\\
        Q221 & How frequently do you vote in local level elections?\\
        Q224 & How often are votes counted fairly in the country's elections?\\
        Q229 & How frequently are election officials fair in country's elections?\\
        Q234 & To what extent do you feel the political system in your country allows people like you to have a say in what the government does?\\
        Q235 & What is your opinion on a political system with a strong leader who does not have to bother with parliament and elections?\\
        Q236 & What is your opinion on a political system that involves experts making decisions based on what they think is best for the country?\\
        Q239 & What is your opinion on a political system governed by religious law with no political parties or elections?\\
        \bottomrule
    \end{tabular}
    \captionof{table}{The set of 31 WVS-7 items considered in this study.}
    \label{tab:wvs-selected-questions-v1}
\end{table*}
 
\section{Selected WVS-7 items}
\label{sec:appendix-wvs}
The World Values Survey (WVS) is a long-running, cross-national survey that collects population-level responses on a broad range of social, cultural, political, ethical, and economic issues using demographically controlled samples worldwide~\cite{haerpfer2022world}. Its seventh wave (WVS-7), conducted between 2017 and 2021, comprises 259 questions administered through localized questionnaires in respondents’ native or dominant languages and includes both globally standardized items and region-specific modules. In this work, we adopt a curated subset of 31 WVS-7 questions introduced in prior work~\cite{alkhamissi2024investigating}, selected to span diverse substantive themes and to probe attitudes that are not purely factual or mechanical. Table~\ref{tab:wvs-selected-questions-v1} presents the full set of selected items. These items are intentionally non-trivial and allow for meaningful variation in responses, making them particularly suitable for evaluating whether persona-conditioned LLMs can reproduce survey-grounded patterns and subgroup differences beyond surface-level agreement.

\begin{table*}[t]
    \begin{minipage}{0.95\textwidth}
        \begin{tcolorbox}[fontupper=\small, coltext=black, width=\textwidth, halign=justify]
            Objective: You are a survey respondent. Answer the following World Values Survey (WVS) question from the point of view of the given persona.\\[2mm]
            Persona (derived from the respondent record):
            \begin{itemize}
            \setlength\itemsep{0em}
                \item Sex: You are \texttt{\{sex\}}
                \item Age and education: You are \texttt{\{age\}} years of age and completed \texttt{\{education\}} education level
                \item Employment: Your current employment status is: \texttt{\{employment\_status\}}
                \item Occupation: Your occupational group is: \texttt{\{occupation\_group\}}
                \item Income: Your household income level is: \texttt{\{income\_level\}}
                \item Religion: When asked about religion, you said: \texttt{\{religion\}}
                \item Ethnicity: Your ethnicity is \texttt{\{ethnicity\}}
            \end{itemize}
            \vspace{1mm}
            
            Instructions:
            \begin{itemize}
            \setlength\itemsep{0em}
                \item Answer strictly from this persona's point of view.
                \item Select exactly one option.
                \item Do not include any commentary.
                \item Return only the number of the chosen option. No words or punctuation.
            \end{itemize}
            \vspace{2mm}
            
            \textbf{Question:} \texttt{\{question\_text\}}\\
            \textbf{Options:}\\
            (1) \texttt{\{option\_1\}}\\
            $\vdots$\\[1mm]
            (K) \texttt{\{option\_K\}}\\[2mm]
            
            Answer:
        \end{tcolorbox}
        \caption{Persona-based (PB) prompt template for non-scale items.}
        \label{tab:wvs-prompt-pb-nonscale}
    \end{minipage}
\end{table*}
 \begin{table*}[t]
    \begin{minipage}{0.95\textwidth}
        \begin{tcolorbox}[fontupper=\small, coltext=black, width=\textwidth, halign=justify]
            Objective: Answer the following World Values Survey (WVS) question.\\[2mm]
            
            Instructions:
            \begin{itemize}
            \setlength\itemsep{0em}
                \item Answer the following question by typing the number corresponding to your chosen answer.
                \item Return only the number of the chosen option. No words or punctuation.
            \end{itemize}
            \vspace{2mm}
            
            \textbf{Question:} \texttt{\{question\_text\}}\\
            \textbf{Options:}\\
            (1) \texttt{\{option\_1\}}\\
            $\vdots$\\[1mm]
            (K) \texttt{\{option\_K\}}\\[2mm]
            
            Answer:
        \end{tcolorbox}
        \caption{Vanilla (V) prompt template for non-scale items.}
        \label{tab:wvs-prompt-v-nonscale}
    \end{minipage}
\end{table*}
 
\section{Prompt Templates, Answer Formatting, and Item Types}
\label{sec:appendix-prompts}
All survey items are instantiated from YAML templates that enforce a consistent prompt structure by standardizing \textit{(i)} the question header, \textit{(ii)} the option header, and \textit{(iii)} the layout of the response options. For non-scale items, options are enumerated as $(1),\ldots,(K)$ and appended verbatim to the prompt. For scale items, we instead use a fixed \enquote{10-point scale} instruction that explicitly anchors the endpoints (``1'' and ``10''). To support controlled prompt-design ablations, the dataset loader provides multiple prompt variants that can be selected via a command-line flag. The active variant is logged alongside each respondent--item instance to enable downstream stratified analyses. For chat-oriented checkpoints, each prompt is wrapped in an instruction-style envelope (i.e., system message followed by the user query). This design ensures that persona text, when included, is confined to the instruction context and does not alter the encoding or ordering of the response options. Tables~\ref{tab:wvs-prompt-pb-nonscale} and~\ref{tab:wvs-prompt-v-nonscale} show the templates of the persona-based and vanilla prompts.

\end{document}